# Assessment of the Shape of an Axisymmetrical Body from Complex Eigenfrequencies

I. G. Efimova

Moscow State University

*Abstract.* Complex eigenfrequencies of the exterior of a scatterer are considered as signatures of the scatterer's shape. A parameter characterizing the ratio of the maximum transverse and longitudinal dimensions of a body is found.

As is known [1, 2], the Prony's method can be applied to obtain the complex eigenfrequencies of the exterior of a body from the body's response to a wideband (video) pulse using. These frequencies are independent of the spatial orientation of an object and the shape of a probing pulse and determined only by the geometry of the object. Therefore, they can serve as signatures of a body's shape. Thus, the imaginary parts of eigenfrequencies can serve for discrimination of axisymmetrical bodies whose ratios of the maximum transverse and longitudinal dimensions $\delta = a/b$ (where *a* is the maximum transverse dimension and *b* is the length of a body along the axis of symmetry, i.e., the longitudinal dimension) do not differ substantially [3]. If bodies noticeably differ in parameter $\delta$, both imaginary and real parts of their complex eigenfrequencies are effective descriptors.

If there is an available technique to extract the pure response of a body from a scattered radar signal, which is usually noisy, the corresponding eigenfrequencies can be calculated from this response and used to classify the body observed. It is supposed that the complex eigenfrequencies of reference objects have been determined preliminarily and summarized in a database. However, when the signatures of an observed object are absent in the database and, therefore, it cannot be classified as one of the reference scatterers, it may be helpful to assess its shape on the whole. For example, certain useful information is carried by aforementioned parameter $\delta$, which characterizes the ratio of the overall dimensions of a body.

Complex eigenfrequencies γα/c (c is the velocity of light in free space) calculated for several axisymmetrical bodies and eigenfrequencies for certain bodies reported in the literature [4, 5] are summarized in the table.

Table. Complex eigenfrequencies (γα/c) of various axisymmetrical bodies

| A body's shape | δ | γα/c |
|---|---|---|
| Sphere [4] | 1 | -0.5 ± i0.86 |
| | | -0.77 ± i1.81 |
| | | -0.83 ± i2.77 |
| | | -0.9 ± i3.7 |
| | | -1 ± i4.63 |
| Truncated cone [3] | 0.749 | -0.29 ± i0.77 |
| | | -0.65 ± i1.65 |
| | | -1.46 ± i2.86 |
| Cone [3] | 0.505 | -0.4 ± i0.58 |
| | | -0.77 ± i2.61 |
| | | -0.94 ± i4.17 |
| Cylinder [3] | 0.5 | -0.54 ± i0.58 |
| | | -0.67 ± i2.04 |
| | | -0.44 ± i3.34 |
| Prolate spheroid [5] | 0.1 | -0.0265 ± i0.1458 |
| | | -0.04 ± i0.2977 |
| | | -0.0497 ± i0.451 |
| | | -0.0582 ± i0.6051 |
| | | -0.0658 ± i0.7598 |
| Thin wire [4] | 0.01 | -0.00129 ± i0.01478 |
| | | -0.00188 ± i0.02957 |
| | | -0.00222 ± i0.04527 |
| | | -0.00259 ± i0.06098 |
| | | -0.0029 ± i0.07577 |

The analysis of the eigenfrequencies presented in the table shows that the imaginary part of the first (in the ascending order of imaginary parts) complex eigenfrequency is an adequate estimate of parameter δ.